\documentstyle[12pt]{article}
\input BoxedEPS
\SetRokickiEPSFSpecial  
\HideDisplacementBoxes

\begin{document}

\begin{titlepage}
\centerline{\Large Signals of Quark Substructure in}
\smallskip
\centerline{\Large Hadron Reactions at Intermediate Energies}
\bigskip

\begin{centering} E.L. Lomon\\[1ex] Center for Theoretical
Physics and\\ Laboratory for Nuclear Science\\ Massachusetts
Institute of Technology\\ Cambridge MA 02139-4307\\
\end{centering}
\bigskip

\centerline{Abstract}\medskip\noindent Exotic hadrons of a few GeV$/\rm c^2$
mass, exhibiting a spectrum
determined by the perturbative interaction of a non-minimal
number of valence quarks, have long been predicted as signals of the
underlying QCD structure.
Realistic models must include short range
(asymptotic freedom) and long range (confinement) effects of QCD\null.
Models that adequately include the confinement region predict exotic
masses 0.2--0.8~GeV higher than the others. The R-matrix
hybrid model has made the most detailed predictions, some of
which have been experimentally observed. This model
postulates a short range valence quark and perturbative gluon
exchange region, connected to a long-range hadronic region by
an R-matrix boundary condition. In this presentation we will give a brief
introduction to the model and then review the status
of its predictions and new applications:

New $Ann$ ($90^\circ$) data in $pp$
elastic scattering corroborates earlier evidence for the lowest mass $I=1$
state,${}^1S_0$ (2.7 GeV$/\rm c^2$).  The lowest predicted $I=0$ di-nucleon,
${}^3S_1$ (2.63~GeV$/\rm c^2$), is consistent with recent
$\Delta\sigma_L(np)$ in that energy region.  Predictions are made
for several other observable di-nucleons with $<3.0$~GeV$/\rm c^2$ mass.

The $S=-2$ di-hyperon sector has long been of interest.
Searches for a bound state (as predicted by some models) have been negative.
The R-matrix calculation predicts a higher mass
$\Lambda\Lambda$--$\Xi N$ resonance at 2.35~GeV$/\rm c^2$. Specific
predictions for the observables in $\Xi N$ scattering are being made.

Spectra for the $\Lambda N$--$\Sigma N$ and the $KN$ exotics
have been predicted.  There is insufficient data at exotic energies, but the
model fits the low energy data well.

Work in progress indicates that the d* (a quasi-bound $\Delta$-$\Delta$)
predicted in a different model is inconsistent with np(${}^3D_3$)
scattering data.  A model of the recently observed d$'$(2065) as a deeply
bound $N$-$S_{11}({}^1S_0)$ is being constructed.

\bigskip
\centerline{\qquad\qquad\qquad MIT-CTP-2680\hfill
nucl-th/9710006\qquad\qquad\qquad}
\end{titlepage}

\section{Introduction}
At very high momentum transfers, where short range effects dominate, the
asymptotic freedom property of QCD correctly predicts that perturbative
quark/gluon processes will dominate particle reaction observations.
Therefore experiments at high energies have been able to identify the SU(3)
color structure, charges and masses of quarks and gluons and the three
families of quarks and leptons, among other properties of QCD\null.  The
other main property expected of QCD, confinement, can be inferred from the
absence of free quarks and gluons in high energy reactions.  However the
specific nature of confinement and the transition to asymptotic freedom can
not be learned from these experiments.  At low energies
we expect, and have overwhelming experimental confirmation, that confinement
of quarks into hadrons is dominant, leading to an effective field theory of
hadrons interacting via exchange of hadrons.  At short distances
between hadrons one again expects quark and gluon degrees of freedom to be
in evidence, but the details of this short range structure can not be
unfolded in the restricted momentum-transfer range at low energies. Only one
parameter, the range of the core, can be determined.  However the low
energy property of this core (it excludes almost all of the wave function
from the inner region) has many possible explanations other than QCD\null.

But at intermediate energies there is an opportunity to learn more about
confinement and its transition to asymptotic freedom.  Not only do the
higher momentum transfers possible allow a finer-grained study of the interior region,
but, in reaching energies corresponding to quark states confined to the
interior, resonance conditions are achieved which enable substantial
penetration of the wave function to the interior.  These resonances carry much
information concerning the perturbative QCD behavior of the multi-quark
configurations (such as cavity corrections to masses) and about the transition
mechanism to confinement (through the width of decay to the simple hadron
components of the multi-quark configuration).  Resonances of this type are
often called ``exotic".  Most of these resonances can be expected to
occur at high enough excitations so that they are very inelastic and in a
background of many partial waves.  This will make them difficult to observe,
but, as we will see, possible to identify by precise measurement of certain
spin observables.  It is possible that some exotic resonances (with
several  heavier quarks in the configuration) may be of low
excitation energy or even bound.  We shall discuss the doubly-strange
dibaryon case which is the best such prospect including only u, d, and s
quarks.

Exotic hadrons of a few
GeV$/\rm c^2$ mass have long been predicted as signals of the
underlying QCD structure. The term ``exotic" refers to bound states or
resonances which are
(i)~constructed from more quarks than the minimal color singlet quark
configurations ($q^3,q\bar q$), of
the ordinary hadrons, and
(ii)~not dominantly a ``molecular" system of quarks already clustered
into the normal hadrons, interacting via hadron exchange.  Exotics would
be characterized by
a spectrum similar to that of the perturbative spectrum of a non-minimal
number of valence quarks.

\begin{figure}[htb]
$$
\BoxedEPSF{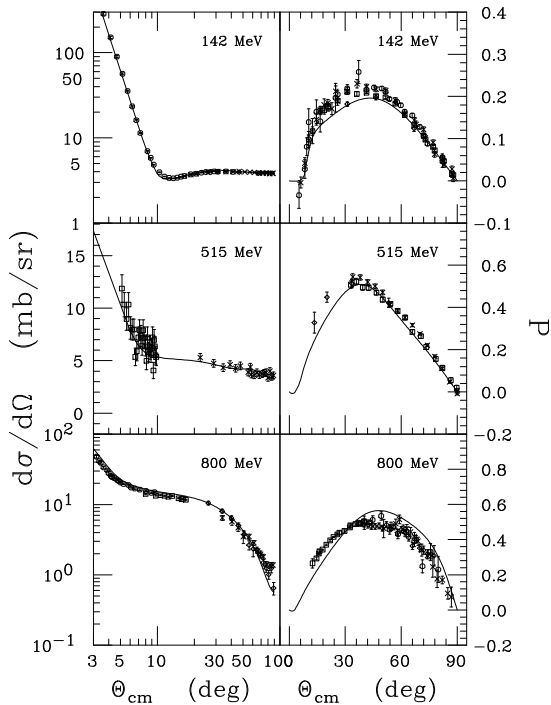 scaled 400}
$$
\caption{{\it Low Energy pp Polarized and Unpolarized Differential Cross Sections\/}: The data
is from the SAID~\cite{SAID} database. The curves are from the R-matrix models of P.~LaFrance,
E.L.~Lomon, and M.~Aw, preprint MIT-CTP\#2133, June 1993.}
\label{fig:tot700}
\end{figure}

A realistic model must enable  a dynamic interaction of the exotic
(asymptotic freedom) and molecular (confinement) effects.
Models that adequately include the confinement region~\cite{LL86,GLL87,MIN88,
CWW82,KNS87} predict masses 0.2--0.8~GeV higher than
those that do not~\cite{AMS78,GOLD89,OY80,RLJ77}. The R-matrix
hybrid model~\cite{LL86,GLL87,MIN88,GL93} has made the most
detailed predictions, some of
which have been experimentally observed. This model
postulates a short range valence quark and perturbative gluon
exchange region, connected to a long-range hadronic region by
an R-matrix boundary condition of the form
\begin{equation}
r_0 \frac{d \psi_\alpha{}^W}{d r_0}
= \sum_\beta f_{\alpha\beta} (W)\ \psi_\beta{}^W (r)
\label{eq:psipeqfpsi}
\end{equation}
where $\psi$ is the 2-hadron external wave function and
$f_{\alpha\beta} (W)$
is a meremorphic function, with real poles of
positive residue
\begin{equation}
f_{\alpha\beta} (W) = f_{\alpha\beta}^0
+ \sum_i \frac{\rho_{\alpha\beta}^i}{W-W_i}
\label{eq:feq}
\end{equation}
with
\begin{equation}
\rho_{\alpha\beta}^i
= -r_0 \frac{\partial W_i}{\partial r_0}
\xi_\alpha^i
\xi_\beta^i
\label{eq:xixi}
\end{equation}
where $W_i$ is the energy of an internal quark state that vanishes at $r_0$,
$i$ running over the complete set with the given quantum numbers,
and the $\xi_\alpha^i$ are the fractional parentage
coefficients (fpc's) of the quark configuration $i$ with the
hadron channel $\alpha$.  The separation radius $r_0$ corresponds to a
distance between the hadron centers-of-mass of $\sim1$~fm, which must satisfy
the R-matrix method condition that the interior description,
perturbative QCD in this case, is a good approximation at
this radius, while simultaneously enabling a fit to the data below the first
exotic.  These conditions are satisfied by the parameters of the Cloudy Bag
Model, but not by those of the MIT Bag Model~\cite{LL86,GLL87}.  This an example
of the greater constraint of hadron reactions on QCD models,
compared to that of single hadron properties.  We display
the fit to the low energy polarized and unpolarized differential pp cross sections, 
see Fig.~\ref{fig:tot700}.

It is important to note that the multi-quark configurations overlap not only
with hadron pairs, but also with colored pairs of quark subsets.  Therefore
the sum of the squares of the fpc's in (\ref{eq:xixi}) is less than unity.
For the case of six S-state quarks (of whatever flavor) the sum is 0.2.  The
result of this "hidden color" is that the resonance widths are only a fraction
of the approximately $r_0^{-1}$ expected if the system simply "fell apart".
This results in widths of the order of 50 MeV, which are discernible
against the
slowly varying background, and yet are easily resolved in medium energy
experiments.

In the following sections we will review the status of several of
the R-matrix model's predictions.

\section{Nucleon-Nucleon Sector: Exotics and Observables} 
Two-baryon systems have six valence quarks and consequently only non-minimal
quark configurations.  In the perturbative limit assumed for exotics, the
lowest mass configurations are the $[q(1s_{1/2}]^6$ even parity
and the $[q(1s_{1/2})]^5 q(1p_{1/2,3/2})$ odd parity cases,
with q either u or d for nucleons.

\begin{figure}[htbp]
$$
\BoxedEPSF{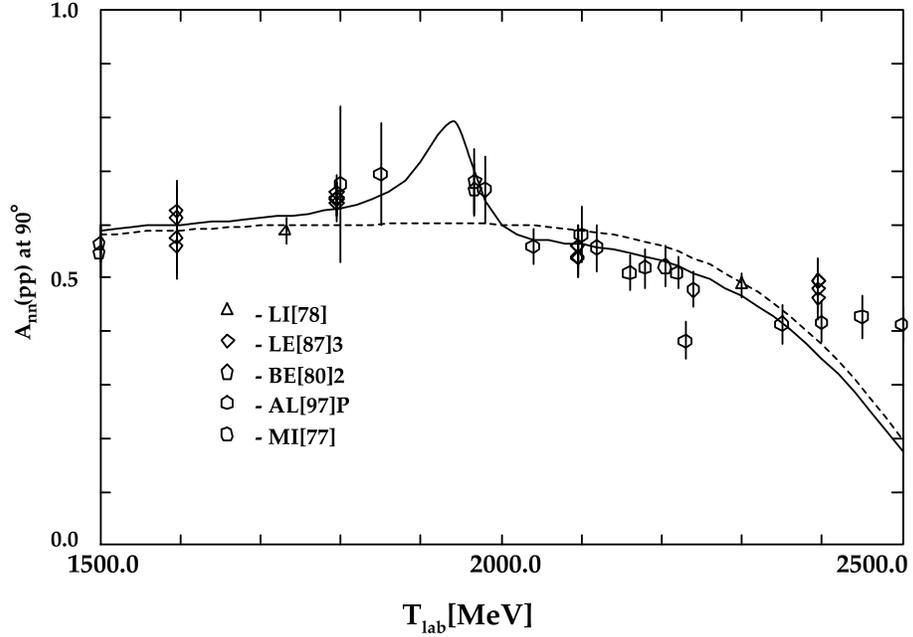 scaled 700}
$$
 \caption{{\it $A_{nn}$(pp) at $90^\circ$}: The data is that of the SAID~\cite{SAID} database
and from the 1997 thesis of C.~Allgower~\cite{SAT93}, a preliminary analysis of 1993--95
Saturne~II data (systematic errors included). The dashed curve is from the SAID
PSA~\cite{SAID} which did not include the Allgower data. The solid curve adds the ${}^1 S_0$
resonance of the 1995 R-matrix model to the SAID PSA\null. The position of the resonance has
been decreased by 20~MeV from the prediction of Ref.~\cite{GLL87}.}
 \label{fig:Ann}
\end{figure}

The lowest mass $I=1$ di-nucleon state, a
${}^1S_0$, which is predicted~\cite{LL86,GLL87} to be at 2.7 GeV$/\rm c^2$,
has been observed in $Ann$ ($90^\circ$) of $pp$ elastic scattering~\cite{SAT93}, see
Fig.~\ref{fig:Ann},  as well as observed~\cite{ZGS89} in $\Delta\sigma_L(pp)$.

\begin{figure}[htbp]
$$
\BoxedEPSF{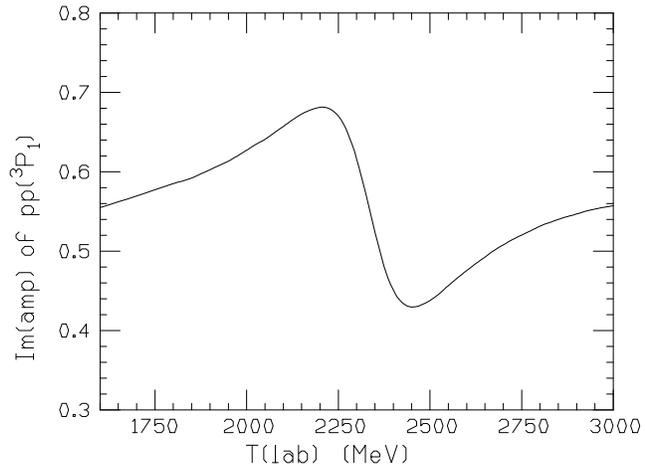 scaled 600}
$$
 \caption{{\it ${}^3P_1$ Exotic Resonance\/}: The pp$({}^3P_1)$ $Im ({\rm amp}) = \frac12
(1-\eta\cos2\delta)$ predicted by the f-pole of the $[q({}^1S_{1/2})]^5 q({}^1 P_{\frac12}]$
configuration~\cite{ELL90}. The c.m.~width is 80~MeV.}
 \label{fig:3P1}
\end{figure}

The lowest mass $I=0$ di-nucleon~\cite{LL86}, a
${}^3S_1$ at (2.63~GeV$/\rm c^2$), is consistent with recent
$\Delta\sigma_L(np)$ data~\cite{DUBNA96} in that energy region. There is also
evidence from $Ay$ in $pp\to d\pi^+$~\cite{BER88} for an $I=1$, spin triplet
di-nucleon near 2.8~GeV$/\rm c^2$) mass. Predictions are made
for several other observable di-nucleons with $\le 3.0$~GeV$/\rm
c^2$ mass~\cite{ELL90}, such as the lowest mass odd parity ${}^3P_1$
resonant structure, Fig.~\ref{fig:3P1}, of about 70 MeV width.

\section{Di-hyperon Sector} 
The $S=-2$ di-hyperon sector has long been of interest
because of an  early prediction of a bound di-lambda (H particle)~\cite{RLJ77}.  The R-matrix
calculation differs, predicting a higher mass
$\Lambda\Lambda-\Xi N$ resonance at 2.35~GeV$/\rm c^2$~\cite{MIN88}.
Searches for the bound state have been negative~\cite{BNL}, leaving the
possibility that the predicted resonance exists.
In fact the exotic in question, a dimF=1, $[q(1s_{1/2})]^4 [s(1s_{1/2})]^2$
quark configuration, overlaps more strongly with the $\Xi$-$N$ and
$\Sigma$-$\Sigma$ channels than with the $\Lambda$-$\Lambda$ channel.
Because the
exotic resonance is predicted to be above the $\Xi$-$N$ threshold, the
effect of
the resonance should be substantial in $\Xi$-$N$ scattering, a process
observable
in the laboratory.  We are constructing a coupled channel R-matrix model for
these channels so that specific predictions can be made for $\Xi$-$N$
scattering, to separate the effects
of the $\Lambda$-$\Lambda$ and $\Sigma$-$\Sigma$ thresholds from that of
the exotic resonance.

\section{Nucleon$-$Hyperon and Kaon$-$Nucleon Sectors}
\begin{figure}[htbp]
$$
\BoxedEPSF{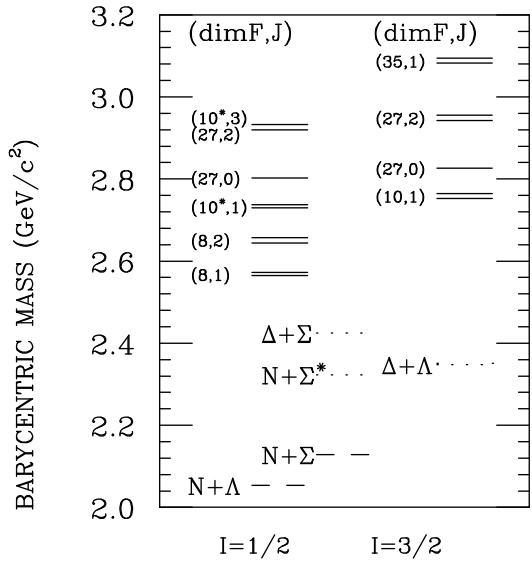 scaled 700}
$$
 \caption{{\it Nucleon-Hyperon Exotic Spectrum\/}: The R-matrix model predictions for the
exotics produced produced by the $[q(1s_{1/2})]^6 [s(1s_{1/2})]$ quark
configurations~\cite{GL93}. The channel thresholds are shown by dashed lines (ground state
baryons) and dotted lines (isobar + ground state). }
 \label{fig:N-Yspec}
\end{figure}

\begin{figure}[htbp]
$$
\BoxedEPSF{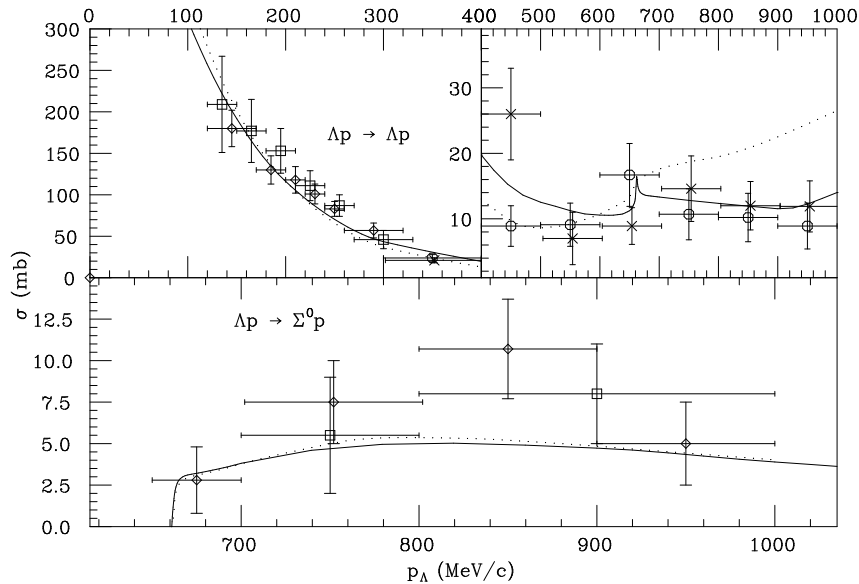 scaled 700}
$$
 \caption{{\it $\Lambda$-$N$ Reactions\/}: The elastic $\Lambda$  total cross section is
shown, with a blow-up of the high energy region to show the effect of the $\Sigma$-$N$
threshold. The $\Lambda p \to \Sigma^0p$ production cross section illustrates the rise from
the threshold. The curves are those of Ref.~\cite{GL93}.}
 \label{fig:LN-SN}
\end{figure}

Spectra for the $\Lambda N - \Sigma N$ and the $KN$ exotics of the
$[q(1s_{1/2})]^5 s(1s_{1/2})$ and the $[q(1s_{1/2})]^4 \bar s$
configurations have
been predicted~\cite{MIN88,GL93}.  The rich exotic nucleon-hyperon spectrum
predicted, together with the relevant thresholds, is shown in Fig.~\ref{fig:N-Yspec}.  Modern
data is not available in the predicted
resonance region, but the R-matrix model can be compared with lower energy
nucleon-hyperon reaction data to which  it is a good fit~\cite{GL93}.  As an
example
we show the total cross sections for $\Lambda$-$N$ elastic scattering and
$\Sigma$-$N$
production, Fig.~\ref{fig:LN-SN}, clearly demonstrating the effect of the
$\Sigma$-$N$
threshold on the elastic $\Lambda$-$N$ prediction, also reflected in the data.

There are hints of structure in old bubble chamber data for the KN system.

\section{Is there a d*?} 
\begin{figure}[htb]
$$
\BoxedEPSF{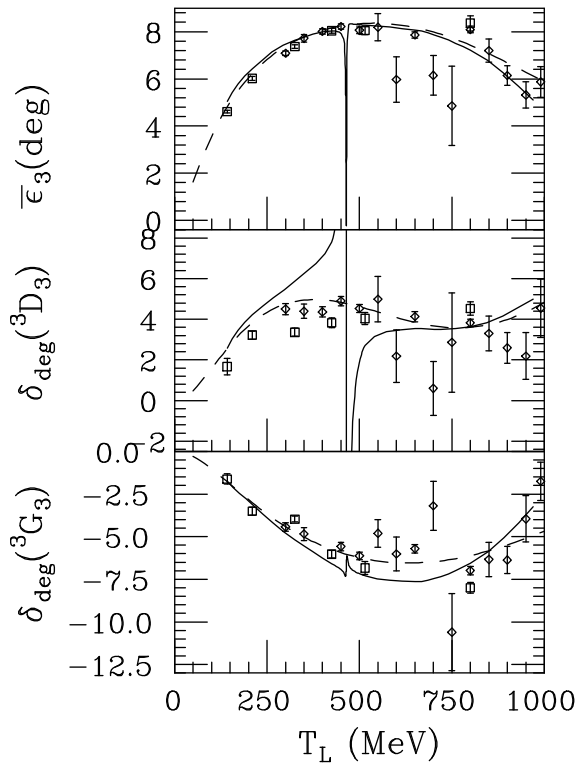 scaled 700}
$$
 \caption{{\it d* Resonance in np$({}^3D_3$-${}^3G_3)$\/}: The coupled channel R-matrix model
is able to predict the effect of a $\Delta$-$\Delta({}^7S_3)$ bound state on the np elastic
scattering parameters. The bound state is induced in the region predicted in
Ref.~\cite{GOLD89} and its width minimized by adjustment of the f-matrix as described in the
text. The fit to the PSA of Refs.~\cite{SAID} and~\cite{BUGG} is very poor near resonance,
especially for $\delta({}^3D_3)$. The best fit of the R-matrix model without a $\Delta$-$\Delta$
bound state (not shown) is very good.}
 \label{fig:3D3}
\end{figure}

In a cluster model which screens the interaction between quarks in different
baryons~\cite{GOLD89} a quasi-bound I=0, ${}^7S_3$ $\Delta$-$\Delta$ state
of mass
near 2.1~GeV$/\rm c^2$ is predicted, but has not yet been observed.  This
channel is coupled to the np(${}^3D_3$-${}^3G_3$) channel by long range tensor
potentials arising from pi- and rho-meson exchange.  It is therefore expected
to have a substantial decay width to that channel.  In np scattering this would
show up as a resonance of equal or greater width at a neutron beam energy
T(lab) = 425-550 MeV.  For $T(lab) \le 800$ MeV there are over-complete sets
of precise spin observables at several energies, in particular at 142, 210,
325, 425, 515, 650 and 800 MeV.  The phase shift analyses~\cite{SAID,BUGG} are
able to determine the ${}^3D_3$-${}^3G_3$ phase angles to 0.4 deg.\ or better,
and produce a smooth energy dependence within that error.  A resonance, elastic
in this energy region, needs to have a width of $\le$ 1 MeV to
reproduce that degree of smoothness.  The R-matrix model with all relevant
$I=0$, $3^{+}$ channels fits the data well, when there is no d*.  It can
produce a
quasi-bound ${}^7S_3$~$\Delta$-$\Delta$ state half way between 425 and 515 MeV
by adjusting the energy-independent
term of that channel's f-matrix component.  Furthermore it can minimize the
width induced by meson exchanges by adjusting the off-diagonal constant
f-matrix
component.  The resultant width is 4 MeV, too large to agree with the phase
shift analyses as seen in Fig.~\ref{fig:3D3}.  There is a greater energy gap
between the well determined phases at T(lab) of 650 and 800~MeV, but the larger
np phase space increases the resonance width.

\section{What is the d$'$?} 
There is considerable evidence indicating a narrow dibaryon resonance at 2.065
GeV, probably in an I=0 state~\cite{d'data}.
It is conjectured to be in a $0^{-}$ state to prevent it from having
a large decay width to an NN channel.  This d$'$ may be a bound $\pi NN$ system~\cite{piNN}. 
The corresponding exotic configuration is more than 0.5~GeV too massive in the R-matrix
model as well as in a constituent quark model calculation~\cite{TUEB}.
Another possibility is that it could be a $N$-$S_{11}$ system
bound in an S-wave by the effect of the constant term in the f-matrix.  We will
explore this with the R-matrix model, which can predict its electromagnetic
decay widths.

\section{Conclusions} 
The R-matrix model of strong interactions incorporates both the long range and
short range properties of QCD permitting the description of two-hadron
interactions that include aspects of both asymptotic freedom and
confinement.
In particular predictions can be made for the spectra of exotic resonances,
while also describing the properties of molecular (or mixed) resonances and
the details of the background reactions.  This enables a better determination
of non-perturbative properties of QCD than obtainable from single hadron
properties alone.
The experimental discovery of more of the exotic spectrum would greatly
refine our knowledge of QCD dynamics.

The R-matrix model assumes a sudden transition from perturbative QCD to the
region of hadron exchange forces.  We are considering the possible introduction
of the chiral symmetric hadron phase between $r_0$ (corresponding to the
deconfinement transition) and the broken symmetry asymptotic hadron region.
\bigskip

\subsection*{Acknowledgments} \medskip
We are grateful to NERSC for its allocation of supercomputer resources, as many
of these calculations required the precision inherent in the Cray word length.
Several of the research ideas came from conversations with A. Faessler, A. Gal, L. Glozman, T.
Goldman 
 and E. Shuryak.  My collaborators in earlier work are
referenced in the cited papers.


\begin{thebibliography}{99}
\frenchspacing
\bibitem{LL86}
P.~LaFrance and E.~Lomon,
{\it Phys.~Rev.\/} {\bf D34}, 1351 (1986).
\bibitem{GLL87}
P.~Gonz\'alez, P.~LaFrance and E.~Lomon,
{\it Phys.~Rev.\/} {\bf D35}, 2142 (1987).
\bibitem{MIN88}
E.~L.~Lomon,
{\it AIP~Conf.~Proc.\/} {\bf 187},
8th~Int.~Symp. on High Energy Physics, K.~J.~Heller ed. (MN, 1988), p. 630
\bibitem{CWW82}
C.~W.~Wong,
{\it Prog.~Part.~Nucl.~Phys.\/} {\bf 8}, 223 (1982);
C.~W.~Wong and K.~F.~Liu,
{\it Phys.~Rev.~Lett.\/} {\bf 41}, 82 (1978).
\bibitem{KNS87}
Yu.~S.~Kalashnikova, I.~M.~Narodatskii and Yu.~A.~Simonov,
{\it Sov.~J. Nucl.~Phys.\/} {\bf 46}, 689 (1987).
\bibitem{AMS78}
A.~Th.~M.~Aerts, P.~J.~G.~Mulders and J.~J.~de~Swart,
{\it Phys.~Rev.\/} {\bf D17}, 260 (1978).
\bibitem{GOLD89}
T.~Goldman, K.~Maltman, G.~J.~Stephenson~Jr., K.~E.~Schmidt and Fan~Wang,
{\it Phys.~Rev.\/} {\bf C39}, 1889 (1989); Fan~Wang et al.,{\it Phys.~\/}
{\bf C51}, 3411 (1995).
\bibitem{OY80}
M.~Oka and K.~Yazaki,
{\it Phys.~Lett.\/} {\bf 90B}, 41 (1980).
\bibitem{RLJ77}
R.~L.~Jaffe,
{\it Phys.~Rev.~Lett.\/} {\bf 38}, 195 (1977).
\bibitem{GL93}
W.~R.~Greenberg and E.~L.~Lomon,
{\it Phys.~Rev.\/} {\bf D47}, 2703 (1993).
\bibitem{SAT93}
J.~Ball et al.,
{\it Phys.~Lett.\/} {\bf 320}, 206 (1994);
C.~Allgower,
{\it Bull.~A.P.S.\/} {\bf 41}, 1266 (1996);
C.~Allgower, thesis 1997.
\bibitem{ZGS89}
I.~P.~Auer et al.,
{\it Phys.~Rev.~Lett.\/} {\bf 62}, 2649 (1989).
\bibitem{DUBNA96}
V.~I.~Sharov et al.,
{\it JINR Rapid Communications\/}, No.3[77] (1996).
\bibitem{BER88}
R.~Bertini et al.,
{\it Phys.~Lett.\/} {\bf 203}, 18 (1988).
\bibitem{ELL90}
E.~Lomon,
{\it J.~Physique\/} {\bf 58} Supple Colloque C6, 363 (1990).
\bibitem{BNL}
R.~W.~Stotzer et al.,
{\it Phys.~Rev.~Lett.\/} {\bf 78}, 3646 {1997}.
\bibitem{SAID}
R.~A.~Arndt et al.,
submitted to{\it Phys.~Rev.\/} {\bf C}.
\bibitem{BUGG}
D.~V.~Bugg,
{\it Phys.~Rev.\/} {\bf C41}, 2708 (1990).
\bibitem{d'data}
W.~Brodowski et al.,
{\it Z.~Phys.\/} {\bf A355}, 5 {1996};
R.~Bilger et al.,
{\it Z.~Phys.\/} {\bf A343}, 491 (1993).
\bibitem{piNN}
A.~Garcilazo,
{\it Phys.~Rev.\/} {\bf C56}, scheduled for no 4, Nov. 1997;
A.~Valcarce, H.~Garcilazo and F.~Fern\'andez,
{\it Phys.~Rev.\/} {\bf C52}, 539 {1995}.
\bibitem{TUEB}
G.~Wagner et al.,
{\it Nucl. Phys.\/} {\bf A594}, 263 (1995).
\end{thebibliography}
\end{document}